\documentclass[pra, aps, twocolumn, groupedaddress, showpacs, superscriptaddress]{revtex4}

\usepackage{slashed}
\usepackage{graphicx}
\usepackage{subfigure}
\usepackage[usenames, dvipsnames]{color}
\usepackage{graphics}
\usepackage{hyperref}
\usepackage{bm}
\usepackage{amsfonts}
\hypersetup{backref,
colorlinks=true,
urlcolor=blue,
linkcolor=blue,
linktoc=page,
citecolor=blue}

\everymath{\displaystyle} 
\begin{document}

\title{Impact of Chiral-Transitions in Quantum Friction}

\author{Muzzamal I. Shaukat}
\affiliation{Institute for Quantum Science and Engineering, Texas A\&M University, College Station, Texas 77843, USA}
\email{muzzamalshaukat@gmail.com}

\author{M\'{a}rio G. Silveirinha}
\affiliation{ Instituto Superior T\'ecnico, University of Lisbon and Instituto de Telecomunica\c{c}\~{o}es, Torre Norte, Av. Rovisco Pais 1,
 Lisbon, Portugal}
\email{mario.silveirinha@tecnico.ulisboa.pt}

\begin{abstract}
We theoretically investigate the role of chiral-transitions in the quantum friction force that acts on a two-level atom that moves with relative velocity $v$ parallel to a planar metallic surface. We find that the friction force has a component that is sensitive to the handedness of the atomic transition dipole moment. In the particular, we show that the friction force can be enhanced by an atomic transition with a dipole moment with a certain handedness, and almost suppressed by the dipole moment with the opposite handedness. Curiously, the handedness of the transition dipole moment that boosts the ground-state friction force is the opposite of what is classically expected from the spin-momentum locking. We explain this discrepancy in terms of the interaction between positive and negative frequency oscillators.
\end{abstract}

\maketitle

\section{Introduction}
Casimir forces have garnered immense theoretical \cite{Zhao2009, Rodriguez2011} and experimental \cite{Munday2007, Bao2010, Sushkov2011, Sushkov2011A} interest, experiencing a remarkable resurgence in recent years. In particular, it has been predicted that fluctuation-induced friction forces can occur between two noncontacting bodies moving relative to each other, even at zero temperature. This peculiar behavior arises due to the influence of quantum fluctuations. \cite{, Pendry1997, Volokitin1998, Volokitin1999, Volokitin2007A, Intravaia2022}. The past decades have seen much work carried out on the investigation of quantum friction
\cite{ Intravaia2022, Pendry2010, Pendry1997, Scheel2009, Zhao2012, Maslovski2013, Silveirinha2022, Intravaia2014, Hoye2015, Milton2016, Intravaia2016, Klatt2016, Lannebere2017, Volokitin1998, Volokitin1999, Volokitin2007A, Silveirinha2014a, Pieplow2015, Maghrebi2013, Intravaia2019, Horsley2012, Silveirinha2013a, Hassani2018, Dedkov2002, Intravaia2015, Klatt2017, Dedkov2017}, which remains a captivating and yet not fully understood phenomenon. For instance, the influence of relativistic effects, non-Markovian dynamics, and unstable regimes where the force fails to reach a steady state are still active areas of research and debate \cite{Intravaia2016b, Klatt2017, Oue2024, Oue2025}.

 A physical picture of quantum friction is that when two bodies are set in a relative motion, the ``image'' dipoles lag behind the original fluctuations, resulting in energy loss. For instance, for an atom moving parallel to a metal surface, the energy transfer can be visualized as the atom inducing dipoles in the metal. These induced dipoles, due to the lack of motion and the dispersive properties of the material, lag behind the atom’s fluctuations. This lag results in a continuous transfer of energy from the atom to the metal slab, which is perceived as energy loss from the atom’s perspective.  The friction force is mostly controlled by the material dispersion and dissipation. 
 
Several approaches have been considered to characterize the friction force  such as the Born-Markov approximation \cite{Scheel2009}, linear-response theory \cite{Dedkov2002} and time dependent perturbation theory \cite{Intravaia2015}. The measurement of quantum friction is extremely challenging in systems with moving bodies, and due to this reason there is still no experimental confirmation of this effect.\par

Chiral quantum optics has emerged as a novel paradigm to tailor the quantum light-matter interactions, providing unprecedented control over the directionality of these interactions  \cite{Lodahl2017}. Chiral emitters coupled to waveguides are the basis for several protocols of quantum networks \cite{Mahmoodian2016} and have been investigated theoretically and experimentally \cite{Pichler2015, Jan2014, Ramos2014}.
The chiral nature of these interactions enables the development of nonreciprocal devices such as circulators \cite{Xia2014, Sayrin2015, Scheucher2016}, optical isolators \cite{Sollner2015}, and quantum gates \cite{Shomroni2014, Pedersen}, opening up exciting possibilities for advanced quantum technologies. Furthermore, the exploration of chirality in the quantum realm can be a powerful tool for optical manipulation \cite{Genet2022}. 

The purpose of this work 
is to elucidate the role of the atomic polarization, and in particular the role of chiral-transitions, in the quantum friction force. To this end, we consider a two-level system (qubit) moving with velocity $v$ parallel to a metallic surface.  Using a quasi-static approximation, we present a detailed study of the impact of the atomic transition dipole-moment on the friction force, discuss the role of surface plasmons and analyze the effect of material dissipation. Interestingly, we find that chiral-type atomic transitions can be the dominant friction mechanism. As  intuitively expected, the handedness of a chiral-type atomic transition can greatly influence its contribution to the friction force due to the strong sensitivity of the coupling with surface plasmons. Intriguingly, we find that the atomic transitions that give the dominant contribution to the friction force have an handeness that is opposite to that expected from classical considerations. We offer an explanation for this property by taking into account that the interactions that lead to friction are associated with non-conserving energy terms.

The article is organized as follows. In Sect. II, we describe the theoretical model of the two-level system that interacts with a metallic surface. In Sect. III, we describe the theoretical formalism and provide general formulas for the friction force in terms of the system Green's function. Then, in Sect. IV we obtain closed analytical expressions for the friction force considering a quasi-static model for the metallic substrate and the weak dissipation limit. In addition, we present a detailed numerical study that highlights the impact of the atomic polarization and of the substrate loss. A summary of the main results is provided in Sect. V. 
\section{Theoretical Model}
\begin{figure}[h!]
\center
\includegraphics[scale=1]{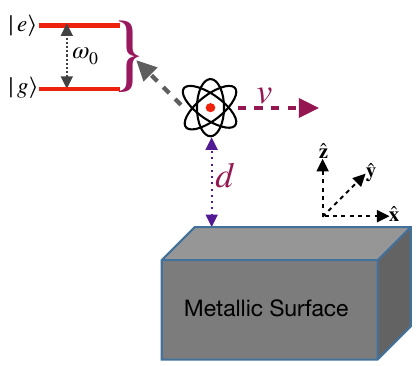}
\caption{A two-level atom moves with velocity $v$ at distance $d$ from a metallic surface.}
\label{Model}
\end{figure}

We consider a scenario where a two-level atom (qubit) travels with velocity $v$ parallel to a thick metallic surface (see Fig. \ref{Model}). The atom moves along the $x$-axis and is at distance $d$ from the metal interface. 
The total Hamiltonian of the system can be written as 
\begin{eqnarray}
\hat{H}=\hat{H}_{\rm at.}+ \hat{H}_{\rm field}+\hat{H}_{\rm int.}, \label{total Hamil.}
\end{eqnarray}
where $\hat{H}_{\rm at.}=\hbar \omega_0 \hat{\sigma}_z/2$ represents the Hamiltonian of the atom. The atomic states are denoted by $\vert e\rangle$ (excited state) and $\vert $g$ \rangle$ (ground state). These states are separated by the energy $E_e -E_g=\hbar\omega_0$, where $\omega_0$ denotes the atomic transition frequency.
Here, $\hat{\sigma}_z$=$\vert e \rangle \langle e \vert - \vert $g$ \rangle \langle $g$ \vert$ denotes the atomic inversion operator. 
The term $\hat{H}_{\rm field}$ represents the Hamiltonian of the plasmonic field, given by
\begin{eqnarray}
\hat{H}_{\rm field}=\sum_{\omega_{n\bold{k}>0}}\frac{\hbar \omega_{n\bold{k}}}{2} \left(\hat{a}_{n\bold{k}}^{\dagger} \hat{a}_{n\bold{k}}+\hat{a}_{n\bold{k}}\hat{a}_{n\bold{k}}^{\dagger}\right),
\end{eqnarray}
where the sum is taken over positive oscillation frequencies $\omega_{n\bold{k}}$, and $\hat{a}_{n\bold{k}}^{\dagger}$ ($\hat{a}_{n\bold{k}}$) denotes the creation (annihilation) operator of the bosonic field, satisfying the commutation relation $[\hat{a}_{n\bold{k}}, \hat{a}_{n\bold{k}'}^{\dagger}]=\delta_{\bold{k},\bold{k}'}$. The term $\hat{H}_{\rm int.} = -\bold{\hat{p}}_d \cdot \bold{\hat{E}}'(\bold{r}_0')$ represents the interaction Hamiltonian between the two level atom and the plasmonic field. Here, ${{{{\bf{r}}}_0'}}$ is the time-independent position of the atom in the co-moving frame. The atom's position in the lab frame is ${{\bf{r}}_0} = {{{\bf{r}}}_0'} + {\bf{v}}t$.

The quantized fields in the lab frame (co-moving with plasmonic slab) can be written in terms of the system electromagnetic modes as \cite{Mario2018, ModalExpansions, Bhat2006, Lannebere2017}
\begin{eqnarray}
\hat{\bold{F}}(\bold{r},t)=\sum_{\omega_{n\bold{k}}>0}\sqrt{\frac{\hbar \omega_{n\bold{k}}}{2}}\left(\hat{a}_{n\bold{k}}(t) \bold{F}_{n\bold{k}}(\bold{r})+ \hat{a}_{n\bold{k}}^{\dagger}(t) \bold{F}_{n\bold{k}}^*(\bold{r})\right), \nonumber \\
\end{eqnarray}
We use a 6-vector notation such that $\bold{F}=(\bold{E} \hspace{2mm} \bold{H})^T$ is formed by the electric and magnetic field components. Strictly speaking, the modal expansion of the quantized field only applies to systems formed by non-dissipative materials. However, as further discussed later in the paper, our derivations can be readily extended to lossy material platforms as the calculated friction force is expressed in terms of the system Green's function.
 
The 6-vector $\bold{F}_{n\bold{k}}(\bold{r})=\bold{f}_{n\bold{k}}(z)e^{i\bold{k}.\bold{r}}$ denotes a generic electromagnetic mode of the metal-air system associated with the frequency $\omega_{n \bf{k}}$. The electromagnetic modes are normalized as explained in Ref. \cite{Mario2018}. Due to the continuous translation symmetry of the system in the $xoy$ plane, the electromagnetic modes have a space dependence in the $x$ and $y$ coordinates of the type: ${e^{i{\bf{k}} \cdot {\bf{r}}}}$ with ${\bf{k}} = {k_x}{\bf{\hat x}} + {k_y}{\bf{\hat y}}
\equiv \left( {k_x ,k_y ,0} \right)$ the in-plane wave vector. The field envelope $\bold{f}_{n\bold{k}}$, depends only on the $z$-coordinate. 

We neglect all the relativistic corrections (Galilean approximation) and, in addition, we identify the electric field operator $\bold{\hat{E}}'$ in the frame co-moving with the atom with the electric field operator $\bold{\hat{E}}$ in the lab frame:  ${\bf{\hat E'}}\left( {{{{\bf{r}}}_0'}} \right) \approx {\bf{\hat E}}\left( {{{\bf{r}}_0}} \right)$. Thus, the magneto-electric coupling arising from the relative motion is neglected. Within these conditions, the interaction Hamiltonian is given by:
\begin{eqnarray}
\hat{H}_{\rm int.}
&\simeq& -\sum_{\omega_{n\bold{k}>0}}\sqrt{\frac{\hbar \omega_{n\bold{k}}}{2}}( \boldsymbol{\tilde{\gamma}}\hat{\sigma}_- +\boldsymbol{\tilde{\gamma}}^* \hat{\sigma}_+) \nonumber \\ &&
\cdot \left(\hat{ a}_{n\bold{k}}\bold{F}_{n\bold{k}}(\bold{r}_0')e^{+i \bold{k}\cdot \bold{v}}+\hat{ a}^{\dagger}_{n\bold{k}}\bold{F}_{n\bold{k}}^*(\bold{r}_0')e^{-i \bold{k} \cdot \bold{v} t} \right).
\label{int. hamiltonian}
\end{eqnarray}
We used $\bold{F}_{n\bold{k}}(\bold{r}_0)=\bold{F}_{n\bold{k}}(\bold{r}_0')e^{+i \bold{k}\cdot \bold{v}}$. 
For convenience, the (generalized) dipole moment operator $\bold{\hat{p}}_d=\boldsymbol{\tilde{\gamma}}^*\hat{\sigma}_+ + \boldsymbol{\tilde{\gamma}}\hat{\sigma}_- $  is defined using a six-vector $\boldsymbol{\tilde{\gamma}}=[\boldsymbol{\gamma} \hspace{0.2cm} 0]^T$, which is written in terms of the standard transition dipole moment (vector) $\boldsymbol{\gamma}$. The atomic raising and lowering operators are defined in a standard way:
 $\hat{\sigma}_+ =\vert e \rangle \langle $g$\vert$ and $\hat{\sigma}_- =\vert $g$ \rangle \langle e\vert$.

\section{Optical Force}
In the electric-dipole approximation, the optical force operator is
\begin{eqnarray}
 \hat{\mathcal{F}}_m =  \bold{\hat{p}}_d \cdot \partial_{m} \bold{\hat{F}}. \hspace{2cm} m=x_0,y_0,z_0 \label{opt. force1}
\end{eqnarray}

Using the Heisenberg equation of motion ($-i\hbar\partial_t A=[H,A]$) and the Born-Markov approximation, one can characterize the dynamics of the inversion operator $\hat{\sigma}_z$ and bosonic operator  $\hat{a}_{n\bold{k}}$,  following the same procedure as in Ref. \cite{Lannebere2017, Mario2018}. Assuming that at initial time the field has no quanta, it is found that:
\begin{eqnarray}
 \mathcal{F}_m &=& \mathcal{P}_e (t) \Xi_1 +\left(1- \mathcal{P}_e (t) \right) \Xi_2 \label{opt. force3}
\end{eqnarray}
where $\mathcal{P}_e(t)$ determines the probability of the excited state  and 
\begin{eqnarray}
\Xi_1 &=&  {\rm Re}  \sum_{\omega_{n\bold{k}>0}}\omega_{n\bold{k}} \frac{  \boldsymbol{\tilde{\gamma}}^* \cdot \partial_m  \bold{F}_{n\bold{k}} \otimes \bold{F}^*_{n\bold{k}} \cdot \boldsymbol{\tilde{\gamma}}} {\omega_{n\bold{k}}' - \omega_0-i 0^+}, \nonumber \\
\Xi_2 &=&  {\rm Re}  \sum_{\omega_{n\bold{k}>0}}\omega_{n\bold{k}} \frac{  \boldsymbol{\tilde{\gamma}} \cdot  \partial_m  \bold{F}_{n\bold{k}} \otimes \bold{F}^*_{n\bold{k}} \cdot \boldsymbol{\tilde{\gamma}}^*} {\omega_{n\bold{k}}' + \omega_0-i 0^+}.
\end{eqnarray}
where $\omega'_{n\bold{k}}=\omega_{n\bold{k}}-\bold{k} \cdot \bold{v}$ is the Doppler shifted frequency in the reference frame of the two-level atom. The electromagnetic modes can be evaluated either at ${\bf{r}}_0$ or at ${\bf{r}}_0'$, as the force is identical in both cases. It should be noted that in the absence of relative motion the force reduces to the result of Ref. \cite{Mario2018}.

We can rewrite Eq. (\ref{opt. force3}) as follows:
\begin{eqnarray}
 \mathcal{F}_m &=&2 \mathcal{P}_e (t)  {\rm Re} \{\boldsymbol{\tilde{\gamma}}^* \cdot \left. {\partial_m\mathcal{G}^+_v} \right|_{\omega  = {\omega _0} + i{0^ + }} \cdot \boldsymbol{\tilde{\gamma}}\} \nonumber \\
&+&2\left(1- \mathcal{P}_e (t) \right) {\rm Re}  \{ \boldsymbol{\tilde{\gamma}}^* \cdot  
\left. {\partial_m \mathcal{G}^-_v} \right|_{\omega  = {\omega _0} + i{0^ + }}
\cdot\boldsymbol{\tilde{\gamma}}\}, \label{friction_modal} 
\end{eqnarray}
with
\begin{eqnarray}
\mathcal{G}^+_v \left(\omega \right) &=&\sum_{\omega_{n\bold{k}}>0} \frac{\omega_{n\bold{k}}}{2}\frac{1}{\omega'_{n\bold{k}}-\omega}\bold{F}_{n\bold{k}} (\bold{r}) \otimes \bold{F}^*_{n\bold{k}} (\bold{r}_0), \nonumber \\
\mathcal{G}^-_v \left(\omega \right)&=&\sum_{\omega_{n\bold{k}}>0} \frac{\omega_{n\bold{k}}}{2}\frac{1}{\omega'_{n\bold{k}}+\omega}\bold{F}^*_{n\bold{k}} (\bold{r}) \otimes \bold{F}_{n\bold{k}} ({\bold{r}}_0).  \label{greena}
\end{eqnarray}

Following Ref. \cite{Lannebere2017}, the time evolution of the probability of the excited state is determined by
\begin{eqnarray}
\mathcal{P}_e(t)&=&\mathcal{P}_e(0) e^{-\left( \Gamma^-+\Gamma^+  \right) t } \nonumber \\
&+&   \frac{\Gamma^-}{\Gamma^-+\Gamma^+}\left(1-e^{-\left( \Gamma^-+\Gamma^+  \right)  t}  \right), 
\label{pe_time}
\end{eqnarray}
where $\mathcal{P}_e(0)$ is the probability of the initial state, and $\Gamma^{\pm}$ are transition rates from the excited (ground) state to the ground (excited) state, respectively. Note that for moving systems $\Gamma^-$ does not need to vanish \cite{Lannebere2017}. Related transition rates for the qubit’s internal dynamics have previously been discussed in Refs. \cite{Klatt2016, Klatt2017}. Note that even for an atom initially prepared in the ground state, the excited state probability approaches a nonzero value, $
{\cal P}_e \left( {t = \infty } \right) = \Gamma ^ -  /\left( {\Gamma ^ -   + \Gamma ^ +  } \right) > 0$, in the steady-state regime. This nonzero value for ${\cal P}_e$ can be viewed as heating of the particle in the nonequilibrium steady state, analogous to a related effect discussed in Ref. \cite{Dedkov2017}.

In Appendices \ref{Positive and negative frequency componenets of the Green's function} and \ref{AppendixGreenFriction} it is shown that the optical force can be written in terms of the system Green's function [Eqs. (\ref{frictionA1})-(\ref{frictionA3})]. Thus, even though the previous derivation assumed that the material is non-dissipative the result given by Eqs. (\ref{frictionA1})-(\ref{frictionA3}) holds true even in dissipative scenarios. The justification is that the response of a lossy system in the upper-half frequency plane (UHP) may be approximated arbitrarily well by the response of a conservative system \cite{Mario2019a, Intravaia2012}. Furthermore, in Appendix \ref{Derivation of Decay Rates}, we derive explicit formulas for the decay rates $\Gamma^{\pm}$ in terms of the system Green's function [Eq. (\ref{decay rates11})].

Our study is focused on the friction force, i.e., on the lateral force component collinear with the direction of the relative motion ($m=x$). It is demonstrated in Appendix \ref{AppendixGreenFriction} that for $v>0$ the friction force can be written explicitly as:

\begin{widetext}
\begin{eqnarray}
\mathcal{F}_x &=&\frac{2\mathcal{P}_e(t) }{2\pi}   \int_{-\omega_0/v}^{\infty} dk_x {\rm Re}\left\{ \boldsymbol{\tilde{\gamma}}^*\cdot ik_x  \mathcal{G}_{k_x}  (\bold{r}_0,\bold{r}_0, \omega_0+k_x v)\cdot \boldsymbol{\tilde{\gamma}}\right\} \nonumber \\
&+& \frac{2(1-\mathcal{P}_e (t))}{2 \pi}   \int_{-\infty}^{-\omega_0/v} dk_x  {\rm Re}\left\{ \boldsymbol{\tilde{\gamma}}^*\cdot ik_x  \mathcal{G}_{k_x}  (\bold{r}_0, \bold{r}_0, \omega_0+k_x v)\cdot \boldsymbol{\tilde{\gamma}}\right\}.
\label{friction force}
\end{eqnarray}
\end{widetext}
In the above, $\mathcal{G}_{k_x}$ is the Fourier transform of the system Green's function (see Appendix B) in the variable $x-x_0$. It is implicit that $\omega_0$ has a small positive imaginary component ($i0^+$). The above formula is totally general apart from the  Born-Markov and the Galilean (non-relativistic) approximations. It generalizes the theory of Ref. \cite{Lannebere2017} to arbitrary dissipative platforms.

In the absence of relative motion ($v=0^+$), the lateral force vanishes identically when the two-level atom is in the ground-state ($\mathcal{P}_e =0$) \cite{Mario2018}. Other restrictions on the lateral force are discussed in Ref. \cite{Hafssaa2019}.

As $\boldsymbol{\tilde{\gamma}}$  is of the form $\boldsymbol{\tilde{\gamma}}=[\boldsymbol{\gamma} \hspace{0.2cm} 0]^T$, the friction force only depends on the $3 \times 3$ sub-array (electric-part) of the Green's function tensor $\mathcal{G}_{EE}$. For a substrate described by the relative permittivity  $\epsilon ({\bf r})$ the Green's function satisfies
\begin{eqnarray}
\nabla  \times \nabla  \times {{\cal G}_{EE}} - {\left( {\frac{\omega }{c}} \right)^2}\varepsilon \left( {\bf{r}} \right){{\cal G}_{EE}} = {\omega ^2}{\mu _0}{{\bf{1}}_{3 \times 3}}\delta \left( {{\bf{r}} - {{\bf{r}}_0}} \right) \nonumber \\
\end{eqnarray}

Following Refs. \cite{Mario2018, Mario2014}, the Green's function in the air region can be decomposed as $\mathcal{G}_{EE}=\mathcal{G}_{EE,0}+\mathcal{G}_{EE,s}$, where the free-space Green's function $\mathcal{G}_{EE,0}=\left(\nabla \nabla +k_0^2 I \right)e^{ik_0 r}/(4 \pi \epsilon_0 r)$ is associated with the self-field, and $\mathcal{G}_{EE,s}$ represents the scattering part of the Green's function. Here, $k_0 = \omega /c$ is the free-space wave number. The tensor $\mathcal{G}_{EE,s}$ is determined by a Sommerfeld's type integral
\begin{eqnarray}
\mathcal{G}_{EE,s} \vert_{z=z_0}  &=& \frac{1}{(2\pi)^2 \epsilon_0}\int\int{dk_x dk_y} e^{i \bold{k}_{\vert \vert} \cdot  \left(\bold{r}-\bold{r}_0 \right)} \nonumber \\
&& \frac{e^{-2\gamma_0 d}}{2\gamma_0} \bold{C}(\omega,\bold{k}_{\vert \vert}).
\label{Sommerfeld Integral}
\end{eqnarray}
In the above, $\gamma_0=-i\sqrt{(\omega/c)^2-\bold{k}_{\vert \vert} \cdot \bold{k}_{\vert \vert}}$ with ${\bold{k}_{\vert \vert}}=k_x \bold{\hat{x}}+k_y \bold{\hat{y}}$ the transverse wave vector, and $d$ the distance of the atom to the interface ($z=0$). The tensor $\bold{C}$ is defined by: 
\begin{eqnarray}
 &&\bold{C}(\omega,\bold{k}_{\vert \vert})=\left[\bold{1}_t+\frac{i}{\gamma_0} \hat{\bold{z}}\otimes \bold{k}_{\vert \vert}  \right]
\cdot   \bold{R}\left( \omega,k_x,k_y \right)\nonumber\\
&&\cdot \left[i\gamma_0  \bold{k}_{\vert \vert} \otimes \hat{\bold{z}} +\left(\frac{\omega}{c} \right)^2 \bold{1}_t-\bold{k}_{\vert \vert} \otimes \bold{k}_{\vert \vert}  \right],
\label{tensorC}
 \end{eqnarray}
where $\bold{1}_t=\hat{\bold{ x}}\otimes\hat{\bold{x}}+\hat{\bold{y}}\otimes\hat{\bold{y}}$ is the transverse identity tensor and $\bold{R}\left( \omega,k_x,k_y \right)$ denotes the $2\times 2$ matrix that links the tangential components of the reflected and incident fields at the substrate interface ($z=0$)
\begin{eqnarray}
   \left( {\begin{array}{cc}
   E_x^{\rm{ref}}  \\
  E_y^{\rm{ref}}  \\
  \end{array} } \right)= \bold{R}(\omega,k_x,k_y)\cdot
  \left( {\begin{array}{cc}
   E_x^{\rm{inc}}  \\
  E_y^{\rm{inc}}  \\
  \end{array} } \right),
\end{eqnarray}
for plane wave incidence from the air region. It is underlined that $\bold{R}$ is defined in the laboratory frame, i.e., in the rest frame of the substrate.

Taking the Fourier transform of Eq. (\ref{Sommerfeld Integral}) in the $x-x_0$ variable and setting ${\bf{r}}={\bf{r}}_0$ in the end one readily finds that
\begin{eqnarray}
\mathcal{G}_{EE,k_x}(\bold{r}_0,\bold{r}_0,\omega) &=& \frac{1}{2\pi \epsilon_0}\int_{-\infty}^{\infty}{ dk_y} \frac{e^{-2\gamma_0 d}}{2\gamma_0} \bold{C}(\omega,\bold{k}_{\vert \vert}). \nonumber \\
\label{Fourier Sommerfeld Integral}
\end{eqnarray}
Note that the contribution of the self-part of the Green's function is disregarded  as it does not play a role in quantum friction. The friction force can be numerically evaluated using the substitutions $\mathcal{G}_{k_x}  \to \mathcal{G}_{EE,k_x} $ and $\boldsymbol{\tilde{\gamma}}  \to \boldsymbol{\gamma}$  in Eq. (\ref{friction force}).


The developed theory is rather general and can be readily applied to calculate the friction force in a wide range of stratified electromagnetic platforms using the system's reflection matrix $\bf{R}$. In particular, it can be used to analyze the friction force in various plasmonic platforms (e.g., metals, graphene, Weyl semimetals) and to study the drag force induced by a drift current \cite{Muzzamal2020, Shapiro, Filipa}, including the effects of spatial dispersion. Related formulations based on the system's Green's function can be found in the literature \cite{Dedkov2017, Intravaia2016, Klatt2017}.
 
\section{Quasi-static approximation}

\subsection{General formalism}

In the following, it is assumed that the atom interacts with a metal surface. 
In the non-relativistic limit, the friction force on the moving two-level atom is mainly determined  by short-wavelength interactions with $k_x \sim \pm \omega_0 /v$, i.e., it is ruled by the short-wavelength surface plasmons. Due to this reason, it is a good approximation to model the interactions of the atom with a metal substrate using a quasi-static formalism.

In a quasi-static description, one can assume that $k_{||} \gg \omega/c$ and thereby $\omega^2/c^2 \rightarrow 0,$ and  $\gamma_0\rightarrow k_{\vert \vert}$. Furthermore, since the surface plasmons have a transverse-magnetic (TM) polarization one can use the result $\bold{R}\rightarrow R \bold{k}_{\vert \vert} \otimes \bold{k}_{\vert \vert}/k^2_{\vert \vert}$. Here, $R$ is the reflection coefficient for TM polarization, which for short-wavelengths can be approximated by:
\begin{equation}
R \approx -\frac{\varepsilon(\omega)-1}{\varepsilon(\omega)+1},
 \label{Rcoef}
\end{equation}
with $\varepsilon(\omega)$ the permittivity of the metal.

In these conditions, the tensor $\bf{C}$ [Eq. (\ref{tensorC})] simplifies to:
\begin{equation}
{\bf{C}} \approx  - R\left( \omega  \right)\left( {{{\bf{k}}_{||}} + i{k_{||}}{\bf{\hat z}}} \right) \otimes \left( {{{\bf{k}}_{||}} - i{k_{||}}{\bf{\hat z}}} \right)
\end{equation}
so that the Green's function $\mathcal{G}_{k_x}$ reduces to
\begin{eqnarray}
\mathcal{G}_{k_x}(\bold{r}_0,\bold{r}_0,\omega) &=& -\frac{1}{2\pi \epsilon_0}\int_{-\infty}^{\infty}{ dk_y}   R\left( \omega \right)\frac{e^{-2 k_{\vert \vert} d}}{2k_{||}}  \nonumber \\ 
&& \left( \bold{k}_{\vert \vert} + i k_{\vert \vert} \hat{\bold{z}} \right) \otimes \left(\bold{k}_{\vert \vert} - i k_{\vert \vert} \hat{\bold{z}} \right).
\label{Fourier Quasi Static}
\end{eqnarray}

Substituting the above formula into Eq. (\ref{friction force}), we obtain the following quasi-static approximation for the friction force ($v>0$):
\begin{widetext}
\begin{eqnarray}
\langle \mathcal{F}_x^{QS} \rangle&=&  \frac{\mathcal{P}_e(t) }{4\pi^2 \epsilon_0}  \int_{-\infty}^{+\infty} dk_y    \int_{-\omega_0/v}^{+\infty}  dk_x  e^{-2k_{\vert \vert}d} k_x A(k_x,k_y)  {\rm Re}\left\{ -iR(\omega_0+k_x v)\right\} \nonumber \\
&+&\frac{(1-\mathcal{P}_e(t)) }{4\pi^2 \epsilon_0}  \int_{-\infty}^{+\infty} dk_y    \int_{-\infty}^{-\omega_0/v}dk_x   e^{-2k_{\vert \vert}d} k_x A(k_x,k_y)  {\rm Re}\left\{ -iR(\omega_0+k_x v)\right\},
 \label{Quasi friction force}
\end{eqnarray}
\end{widetext}
where $A$ is the polarization dependent factor:
\begin{eqnarray}
A\left( {{k_x},{k_y}} \right) = \frac{1}{{{k_{||}}}}{\left| {\left( {{{\bf{k}}_{||}} - i{k_{||}}{\bf{\hat z}}} \right) \cdot {\boldsymbol{\gamma }}} \right|^2}.
\label{Apolfactor}
\end{eqnarray}
It can be checked that for a qubit with a linearly polarized transition dipole moment and $\mathcal{P}_e = 0$,   Eq. (\ref{Quasi friction force}) coincides with the result reported by Intravaia \emph{et al} obtained using the fluctuation-dissipation theorem in \cite{Intravaia2016b} (see their Eqs. (33)-(34)). Moreover, for a vertical dipole it is also consistent with the findings of Dedkov \emph{et al} in \cite{Dedkov2017} (see their Eq. (78)), apart from a multiplicative factor, provided the standard semiclassical model for atomic polarizability is adopted. 

It is important to note that the Born-Markov approximation, used in the derivation of Eq. (\ref{Quasi friction force}), may fail to accurately describe quantum friction across a wide range of non-relativistic velocities \cite{Klatt2017, Jentschura2015}. Specifically, the Born-Markov result is only a valid approximation for the friction force when the kinetic energy is comparable to (or greater than) the atomic and material transition energies. This condition establishes a lower velocity bound below which the theory becomes invalid. This issue will be discussed further in Sect. \ref{Sec:LinPol}.

We shall suppose that the metal permittivity is of  the form $\varepsilon(\omega)=1-2\omega^2_{\rm sp}/\omega(\omega+i \Gamma_c)$, where $\omega_{\rm sp}$ denotes the surface plasmon resonance and $\Gamma_c$ represents the damping (collision rate). In this case, it is possible to write the reflection coefficient $R$ [Eq. (\ref{Rcoef})] as
\begin{eqnarray}
R\left( \omega  \right) =  - \frac{{\omega _{{\rm{sp}}}^2}}{{2{{\omega}_{{\rm{sp}}}'}}}\underbrace {\left[ {\frac{1}{{{{\omega}_{{\rm{sp}}}'} - i\Gamma_c /2 - \omega }} + \frac{1}{{{{\omega}_{{\rm{sp}}}'} + i\Gamma_c /2 + \omega }}} \right]}_{g\left( \omega  \right)} \nonumber \\
\label{Rcoef2}
\end{eqnarray}
where $\omega_{\rm sp}'=\sqrt{\omega_{\rm sp}^2-(\Gamma_c/2)^2}$.

\subsection{Weak dissipation limit}

Next, we consider the weak dissipation limit such that  $\Gamma_c \rightarrow 0^+$. Note that the friction force does not need to vanish in this limit, as there are decay channels due to the emission of plasmons, i.e., due to radiation emission.

In the $\Gamma_c \rightarrow 0^+$ limit, the $g$ function in Eq. (\ref{Rcoef2}) is such that:
\begin{eqnarray}{\mathop{\rm Im}\nolimits} \left\{ {g\left( \omega  \right)} \right\} = \pi \delta \left( {\omega _{{\rm{sp}}}^{} - \omega } \right) - \pi \delta \left( {\omega _{{\rm{sp}}}^{} + \omega } \right).
\end{eqnarray}
Therefore, the friction force is controlled by plasmons with wave number $k_x$ such that ${\omega _{{\rm{sp}}}}  \mp \left( {{\omega _0} + {k_x}v} \right) = 0$, which yields the selection rule: ${k_x} = {k_{{\rm{P}}, \pm }} \equiv  - \left( {{\omega _0} \mp {\omega _{{\rm{sp}}}}} \right)/v$. In particular, the integration in $k_x$ in in Eq. (\ref{Quasi friction force}) can be done analytically as it depends only on the poles of the reflection coefficient $k_{{\rm{P}}, \pm }$:
\begin{eqnarray}
\langle \mathcal{F}_x^{QS} \rangle \vert_{\Gamma_c=0^+}&=& - \frac{\mathcal{P}_e (t) }{8\pi \epsilon_0} \frac{\omega_{\rm sp}}{v} \int_{-\infty}^{+\infty} dk_y  \nonumber \\
&&  \left[ k_x e^{-2k_{\vert \vert}d}  A(k_x,k_y) \right] _{k_x=k_{\rm{P},+}}
 \nonumber \\
&+&\frac{(1-\mathcal{P}_e (t)) }{8\pi \epsilon_0} \frac{\omega_{\rm sp}}{v} \int_{-\infty}^{+\infty} dk_y    \nonumber \\
&&   \left[ k_x e^{-2k_{\vert \vert}d}  A(k_x,k_y) \right]_{k_x=k_{\rm{P},-}}.
\end{eqnarray}

Similarly, it is possible to evaluate the decay rates $\Gamma^{\pm}$ [Eq. (\ref{decay rates2})] that control the time evolution of the atomic state using the quasi-static approximation. A straightforward analysis shows that:
\begin{eqnarray}
{\Gamma ^ \pm } = \frac{1}{{8\pi {\varepsilon _0}\hbar }}\frac{{\omega _{{\rm{sp}}}^{}}}{v}\int\limits_{ - \infty }^{ + \infty } {d{k_y}} {\left[ {{e^{ - {k_{||}}2d}}A\left( {{k_x},{k_y}} \right)} \right]_{{k_x} = {k_{{\rm{P}}, \pm }}}}.
\end{eqnarray}
The friction force can be expressed in terms of the decay rates as follows:
\begin{eqnarray}
\langle \mathcal{F}_x^{QS} \rangle \vert_{\Gamma_c=0^+} &=& {\mathcal{P}_e (t)} \hbar \left( {\frac{{{\omega _0} - {\omega _{{\rm{sp}}}}}}{v}} \right){\Gamma ^ + } \nonumber \\
&&- \left( {1 - {\mathcal{P}_e (t)}} \right)\hbar \left( {\frac{{{\omega _0} + {\omega _{{\rm{sp}}}}}}{v}} \right){\Gamma ^ - }. \nonumber \\
\label{friction_lossless}
\end{eqnarray}
Furthermore, from Eq. (\ref{pe_time}) one sees that in the $t \to \infty$ limit $\mathcal{P}_e (\infty) = \Gamma^-/(\Gamma^-+\Gamma^+)$. Thus, the friction force after the atom reaches a steady-state is:
\begin{eqnarray}
{\left. {\langle \mathcal{F}_x^{QS} \rangle} \right|_{\scriptstyle \Gamma_c=0^+ \hfill\atop
\scriptstyle t \to \infty \hfill}}
= 
 - \frac{{2\hbar {\omega _{{\rm{sp}}}}}}{v}\frac{{{\Gamma ^ - }{\Gamma ^ + }}}{{{\Gamma ^ - } + {\Gamma ^ + }}}.
 \end{eqnarray}
 As the two decay rates ($\Gamma^{\pm}$) are strictly positive the force is negative, i.e., it acts to slow down the relative motion as it should be. Furthermore, it can also be checked that when the atom is prepared in the ground state the force is always negative. Note that in all the previous formulas it is implicit that $v>0$.

 A little analysis shows that the integrals for the decay rates can be written as:
 \begin{widetext}
 \begin{eqnarray}
{\Gamma ^ \pm } = \frac{1}{{8\pi {\varepsilon _0}\hbar }}\frac{{{\omega _{{\rm{sp}}}}}}{v}\int\limits_{ - \infty }^{ + \infty } {d{k_y}} {e^{ - {k_{||}}2d}}{\left[ {\frac{{k_x^2}}{{{k_{||}}}}\left( {{{\left| {{\gamma _x}} \right|}^2} - {{\left| {{\gamma _y}} \right|}^2}} \right) + {k_{||}}\left( {{{\left| {{\gamma _z}} \right|}^2} + {{\left| {{\gamma _y}} \right|}^2}} \right) - i{k_x}\left( {{\boldsymbol{\gamma }} \times {{\boldsymbol{\gamma }}^*}} \right) \cdot {\bf{\hat y}}} \right]_{{k_x} = {k_{{\rm{P,}} \pm }}}},
\end{eqnarray}
where ${\boldsymbol{\gamma }} = \left( {{\gamma _x},{\gamma _y},{\gamma _z}} \right)$. The above integral can be analytically evaluated in terms of modified Bessel functions of the second kind $K_n$ of order $n$ as follows:
\begin{eqnarray}
{\Gamma ^ \pm } &=& \frac{1}{{4\pi {\varepsilon _0}\hbar }}\frac{{\omega _{{\rm{sp}}}^{}}}{v}k_{{\rm{P}}, \pm }^2\left\{ {\left( {{{\left| {{\gamma _x}} \right|}^2} - {{\left| {{\gamma _y}} \right|}^2}} \right){K_0}\left( {2\left| {{k_{{\rm{P}}, \pm }}} \right|d} \right) + } \right.\frac{{{{\left| {{\gamma _z}} \right|}^2} + {{\left| {{\gamma _y}} \right|}^2}}}{2}\left[ {{K_0}\left( {2\left| {{k_{{\rm{P}}, \pm }}} \right|d} \right) + {K_2}\left( {2\left| {{k_{{\rm{P}}, \pm }}} \right|d} \right)} \right] \nonumber \\
&& \left. { - i\,{\mathop{\rm sgn}} \left( {{k_{{\rm{P}}, \pm }}} \right)\left( {{\boldsymbol{\gamma }} \times {{\boldsymbol{\gamma }}^*}} \right) \cdot {\bf{\hat y}}\;{K_1}\left( {2\left| {{k_{{\rm{P}}, \pm }}} \right|d} \right)} \right\}.
\label{Gamma_lossless}
\end{eqnarray}
 \end{widetext}
 The contribution of the terms ${\left| {{\gamma _m}} \right|^2}$ ($m=x,y,z$) is strictly positive. On the other hand, the contribution of the term in the second line of the formula can be positive or negative depending on the handedness of the dipole polarization, i.e., on the projection of the spin angular momentum along $y$-direction. For very small velocities or large distances the decay rates are ruled by the exponential decay of the modified Bessel functions, and thereby are exponentially weak. On the other hand, for small velocities, the rate $\Gamma^ +$ is boosted when $\omega_0 \approx \omega_{\rm{sp}}$, i.e., when the detuning of the atomic transition frequency relative to the plasmon resonance is small.

\subsection{Linearly polarized transitions}
\label{Sec:LinPol}

\begin{figure}[h!]
\center
\includegraphics[scale=0.6]{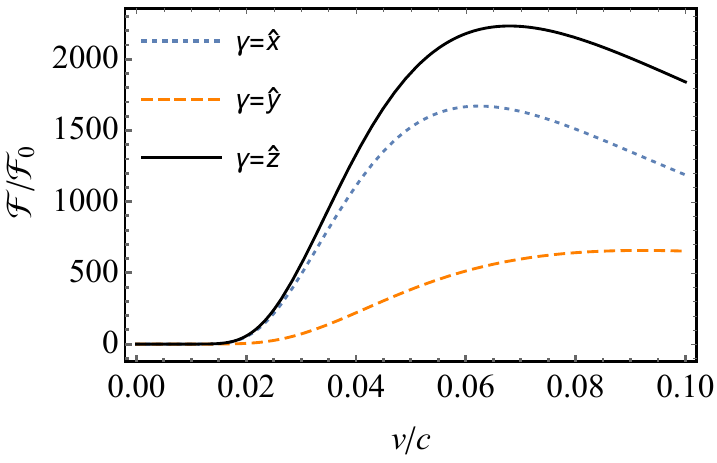}
\caption{Normalized friction force as a function of the velocity for a linearly polarized atom. The atom is prepared in the ground state ($\mathcal{P}_e(t)=0$). The dissipation in the metal is vanishingly weak ($\Gamma_c \rightarrow 0^+$). The atomic transition frequency is $\omega_0=0.1\omega_{\rm{sp}}$ and $d= 0.1 c/\omega_{\rm{sp}}$.}
\label{FigLinearPol}
\end{figure}

 In order to illustrate the ideas, first we consider the case of a linearly polarized atom. Figure \ref{FigLinearPol} depicts the calculated friction force as a function of $v$ for an atom initially prepared in the ground state ($\mathcal{P}_e=0$), for different orientations of the polarization $\boldsymbol{\gamma}$. The force is normalized to  $\mathcal{F}_0 =  - \frac{{{{\left| {\bf{\gamma }} \right|}^2}}}{{4\pi {\varepsilon _0}}}{\left( {\frac{{\omega _{{\rm{sp}}}^{}}}{c}} \right)^4}$. 
 The distance between the atom and the metal surface is taken equal to $d=0.1 c/\omega_{\rm{sp}}$. For the case of silver one can estimate that $\omega_{\rm{sp}}/2\pi= 646{\rm THz}$, which corresponds roughly to  $d=7.4 {\rm nm}$. As seen Fig. \ref{FigLinearPol}, the force tends to increase with the relative velocity $v$, and is always stronger for the case of vertical polarization $ \boldsymbol{\gamma} \sim {{\bf{\hat z}}}$. The weakest force is obtained when the atom polarization is perpendicular to the plane of motion $ \boldsymbol{\gamma} \sim {{\bf{\hat y}}}$. 

Using the asymptotic expansion of the modified Bessel functions for large arguments, it is simple to check that for small $v$ the ground-state friction force depends on the velocity as 
${{\mathcal{F}}_x^{QS} } \sim {u^{7/2}}{e^{ - 2u}}$ with $u=\left| {{k_{{\rm{P}} - }}} \right|d$.
In particular, the friction force becomes exponentially small in the limit $v \to 0$, as illustrated in Fig. \ref{FigLinearPol}. This behavior is consistent with the results reported in previous works \cite{Intravaia2016b, Klatt2017} for calculations up to second order in the atom-field coupling ($\boldsymbol{\gamma}$). As already discussed, the Born-Markov approximation can be problematic for very small values of the velocity \cite{Jentschura2015, Klatt2017}. The lower velocity limit can be estimated as $v_{\rm{min}} \sim w_0 d$, which, for the plots in Fig. \ref{FigLinearPol}, is on the order of $v_{\rm{min}} = 0.01 c$. For smaller velocities, fourth-order interactions contribute an additional term to the force, which leads to an algebraic dependence of the friction force on the velocity, the exact form of which remains a subject of debate \cite{Klatt2017}.

The function ${u^{7/2}}{e^{ - 2u}}$ reaches the maximum at $u=7/4$. Thus, one can estimate that the velocity that maximizes the friction force is on the order of $v_{\rm{opt}} = \frac{4}{7}\left( {{\omega _0} + {\omega _{{\rm{sp}}}}} \right)d$ (here, it is implicit that ${\omega _{{\rm{sp}}}} d \ll 1$). This estimate is consistent with the friction force peaks in Fig. \ref{FigLinearPol}. Furthermore, an identical estimate was derived in Ref. \cite{Intravaia2016b} using second-order perturbation theory. The approach in Ref. \cite{Intravaia2016b} also provides explicit analytical expressions for the friction force in terms of modified Bessel functions. However, unlike our study, it does not account for the possibility of chiral-dipolar transitions. 

\begin{figure}[h!]
\center
\includegraphics[scale=0.65]{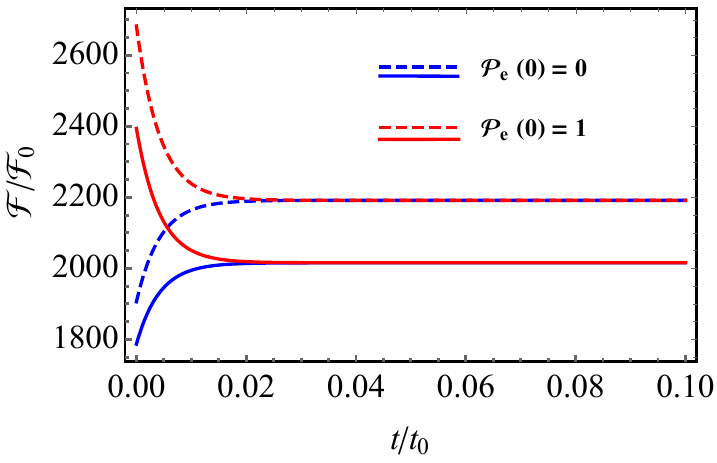}
\caption{Time evolution of the normalized friction force for an atom with $\boldsymbol{\gamma }\sim \bf{\hat z}$ and $v=0.05 c$ initially prepared either in the excited state (red lines) or in the ground state (blue lines). The time normalization factor is $t_0 = 1/\Gamma_0$ with ${\Gamma _0} = \frac{1}{{4\pi {\varepsilon _0}\hbar }}{\left( {\frac{{{\omega _{{\rm{sp}}}}}}{c}} \right)^3}{\left| {\bf{\gamma }} \right|^2}$. Dashed lines: substrate with vanishingly small dissipation. Solid lines: lossy substrate with  $\Gamma_c =0.2\omega_{\rm{sp}}$.}
\label{Timevariation}
\end{figure}

Figure \ref{Timevariation} depicts the time evolution of the normalized friction force when the atom is either initially prepared in the ground state (red curves) or in the excited state (blue curves) for the velocity $v=0.05c$.
The atom is vertically polarized ($\boldsymbol{\gamma }\sim \bf{\hat z}$) and the remaining structural parameters are as in the previous example. The dashed curves are calculated with vanishingly small material dissipation, whereas the solid curves consider that $\Gamma_c =0.2\omega_{\rm{sp}}$ (see section \ref{subsectdiss} for the detailed discussion of the effect of loss). As seen, after some transient determined by the total decay rate (${\Gamma ^ + } + {\Gamma ^ - }$), the atom configuration reaches a steady-state and the friction force becomes constant. For sufficiently large velocities, the excited state probability for $t \to \infty$ approaches ${1 \mathord{\left/
 {\vphantom {1 2}} \right.
 \kern-\nulldelimiterspace} 2}$ (not shown). 
In the present example, the friction force in the excited state is larger than in the ground state. It should be noted that the sign of the lateral force when the atom is in the excited state is typically unconstrained, i.e., it does not need to be anti-parallel to the direction of relative motion \cite{Scheel2009, Mario2018}. Furthermore, It is also worth noting that for a vertical dipole the lateral force vanishes in the limit $v \to 0^+$, independent of the initial atomic state. This is so because the radiation pattern of the dipole has a continuous rotation symmetry with respect to the $z$-direction \cite{Mario2018}.

\subsection{Chiral-type transitions}

\begin{figure}[h!]
\center
\includegraphics[scale=0.6]{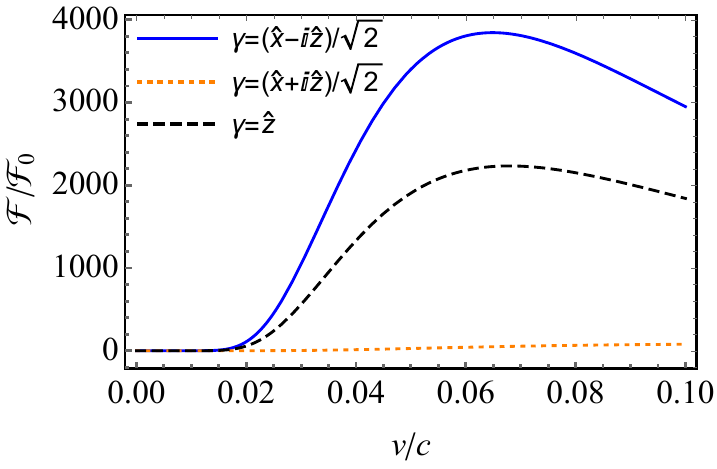}
\caption{Similar to Fig. \ref{FigLinearPol} but for an atom with a chiral-type transition dipole moment.}
\label{FigCircPol}
\end{figure}

Next, we investigate how chiral-type polarized states can affect the friction force (Fig. \ref{FigCircPol}). To this end, we  consider a circularly polarized atom with a transition dipole moment $\boldsymbol{\gamma }\sim \frac{1}{{\sqrt 2 }}\left( {{\bf{\hat x}} \pm i{\bf{\hat z}}} \right)$. Note that the polarization curve is contained in the plane of motion ($xoz$ plane), as that is configuration that facilitates the interaction with plasmons. For comparison, the figure also depicts the force associated with vertical polarization. As seen, the chiral-polarized states have a dramatic effect on the strength of the friction force. In particular, the force is enhanced for a transition dipole moment of the type $\boldsymbol{\gamma }\sim \frac{1}{{\sqrt 2 }}\left( {{\bf{\hat x}} - i{\bf{\hat z}}} \right)$, whereas it is nearly suppressed for a dipole moment with the opposite handedness $\boldsymbol{\gamma }\sim \frac{1}{{\sqrt 2 }}\left( {{\bf{\hat x}} + i{\bf{\hat z}}} \right)$. The different behavior is due to the spin-dependent term (proportional to ${\left( {{\boldsymbol{\gamma }} \times {{\boldsymbol{\gamma }}^*}} \right) \cdot {\bf{\hat y}}}$) in Eq. (\ref{Gamma_lossless}). It can be shown that the handedness of the plasmons that co-propagate with the two-level atom ($+x$-direction) in the air region is such that the corresponding electric field satisfies ${\bf{E}}\sim{\bf{\hat x}} + i{\bf{\hat z}}$. Heuristically, from the spin-momentum locking \cite{Bliokh,Jacob}, one might expect that the friction force should be boosted when ${\bf{E}}$ matches $\boldsymbol{\gamma}$, in contradiction with the results of Fig. \ref{FigCircPol}. 

The different behavior can be justified as follows. When the atom velocity vanishes, the light-matter interactions are mainly determined by  energy-conserving terms of the type $\hat a_{n{\bf{k}}} {\hat \sigma _ + }$. Such terms can lead to a resonant coupling because the field and atomic operators have frequencies with opposite signs (counter-rotating terms). Thus, in the absence of motion the strength of the resonant interactions are controlled by the overlap term ${{\bf{E}}_{n{\bf{k}}}} \cdot {{\bf{\gamma }}^*}$. In contrast, when the relative velocity is nontrivial, there are energy non-conserving terms of the type $\hat a_{n{\bf{k}}}^\dag  {\hat \sigma _ + }$, which can be resonant due to the Doppler shift suffered by the atomic frequency. Specifically, this can happen when the Doppler-shifted frequency $\omega_{0}+k_x v$ becomes $negative$ and matches the plasmon resonance $- \omega_{\rm{sp}}$. The energy non-conserving terms are the ones that activate the friction force when the system is initially prepared in the ground state. The strength of the corresponding interactions is determined by the overlap term ${\bf{E}}_{n{\bf{k}}}^* \cdot {{\bf{\gamma }}^*}$. The two overlap terms discussed previously (${\bf{E}}_{n{\bf{k}}} \cdot {{\bf{\gamma }}^*}$ and ${\bf{E}}_{n{\bf{k}}}^* \cdot {{\bf{\gamma }}^*}$) are evidently maximized for an opposite handedness of the transition dipole moment. This property justifies why the dipole handedness that maximizes the ground state friction force is opposite to the handedness that maximizes the classical light matter interactions in the absence of relative motion.

From a different perspective, one can observe that in the friction problem the interaction between the atom and the substrate leads to the emission of a plasmon with wave number $k_x = k_{\rm{P},-}<0$, oscillation frequency $\omega_{0}+k_x v = - \omega_{\rm{sp}}$ and polarization $\boldsymbol{\gamma}$. Due to the reality of the electromagnetic field, the plasmon with the negative oscillation frequency is equivalent to a plasmon with positive oscillation frequency $\omega_{\rm{sp}}$ but with polarization $\boldsymbol{\gamma}^*$. Thus, when the interaction is mediated by \emph{negative} frequencies, the resonant interaction between the plasmons and the atom corresponds to the matching of  ${\bf{E}}\sim{\bf{\hat x}} + i{\bf{\hat z}}$ with $\boldsymbol{\gamma}^{*}$. This property justifies why the polarization that boosts the ground-state friction force satisfies $\boldsymbol{\gamma }\sim \frac{1}{{\sqrt 2 }}\left( {{\bf{\hat x}} - i{\bf{\hat z}}} \right)$.

\begin{figure}[h!]
\center
\includegraphics[scale=0.65]{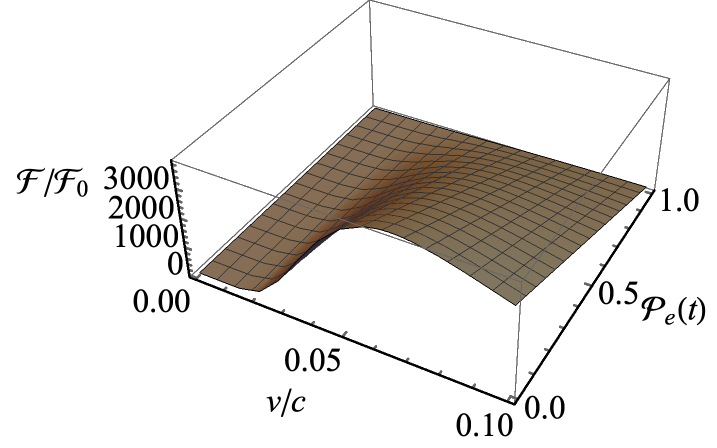}
\caption{Normalized friction force for circular polarization ($\boldsymbol{\gamma }\sim \frac{1}{{\sqrt 2 }}\left( {{\bf{\hat x}} - i{\bf{\hat z}}} \right)$)  as a function of the velocity and atom configuration. The structural parameters are as in Fig. \ref{FigCircPol}.}
\label{Polarized without losses3D}
\end{figure}

Figure \ref{Polarized without losses3D} depicts a 3D plot of the friction force calculated for $\boldsymbol{\gamma }\sim \frac{1}{{\sqrt 2 }}\left( {{\bf{\hat x}} - i{\bf{\hat z}}} \right)$ as a function of the velocity and $\mathcal{P}_e$. As seen, the force is particularly strong when the atom is in the ground state, and is almost suppressed when the atom is in the excited state. This result is due to the fact that the interactions with the excited state are mediated by \emph{positive} frequencies and thus are boosted for an electric dipole with the opposite handedness, consistent with the previous discussion.

It is relevant to note that the impact of spin-momentum locking on the frictional force has been previously discussed in the context of quantum interactions between a Lorentz quantum harmonic oscillator and a metal surface \cite{Intravaia2019}. It was found that the inclusion of rotational degrees of freedom leads to a reduction in the friction force and induces a rotational motion of the atom (“quantum rolling”), which occurs in a direction opposite to that expected in classical mechanics.

For time-reversal-invariant qubit systems, the presence of a chiral transition with a specific handedness necessarily implies the existence of a corresponding transition with the opposite handedness \cite{Mario2019}. In such scenarios, a more accurate representation of a spin-integer qubit is a V-type qubit with two degenerate excited states connected by time-reversal symmetry. The electric dipole moments for transitions between the ground and excited states are related by complex conjugation. As a result, a time-reversal-invariant qubit typically exhibits effects associated with both chiralities. In qubits with a V-type structure, the most significant chiral transition is the one associated with the largest frictional force. As previously noted, this transition corresponds to the handedness opposite to what might be intuitively expected from spin-momentum locking.

Finally, it is important to emphasize that a chiral qubit with a V-type energy structure does not need to exhibit electromagnetic chirality in the conventional sense. This is because it lacks a magneto-electric coupling or a net gyrotropic response due to the coexistence of both chiralities.

\subsection{Effect of dissipation \label{subsectdiss}}

\begin{figure*}
\begin{widetext}
  \centering
   \subfigure[\label{fig:a}]{{\includegraphics[scale=0.67]{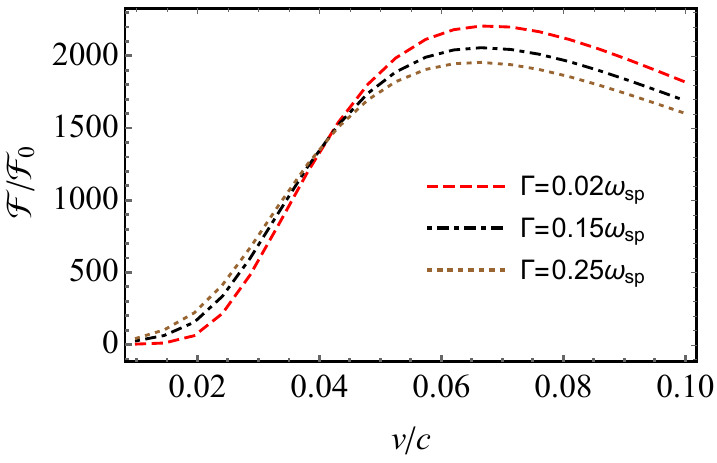}}}
 \subfigure[\label{fig:b}]{{\includegraphics[scale=0.66]{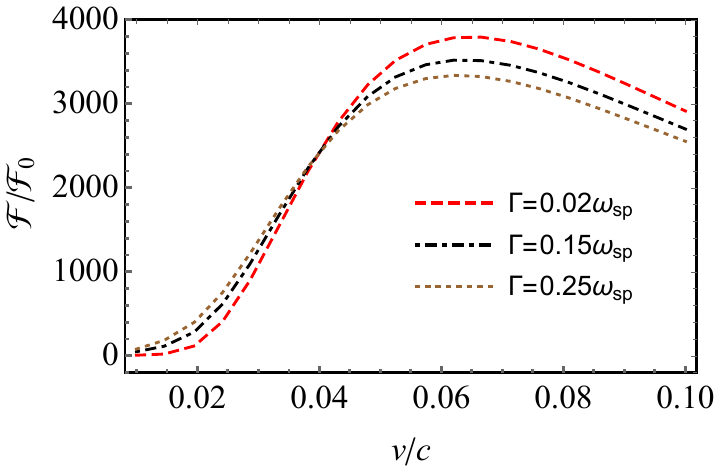}}}
    \caption{Effect of dissipation on the ground-state ($\mathcal{P}_e(t)=0$) friction force. (a) Vertical transition dipole-moment with $\boldsymbol{\gamma }\sim {\bf{\hat z}}$. (b) Chiral-type transition dipole-moment with $\boldsymbol{\gamma }\sim \frac{1}{{\sqrt 2 }}\left( {{\bf{\hat x}} - i{\bf{\hat z}}} \right)$.
    The transition frequency is  $\omega_0=0.1\omega_{sp}$, $d=0.1 c/{\omega_{\rm{sp}}}$.}
    \label{Figlosses}
 \end{widetext}
\end{figure*}

It is relevant to investigate the impact of material dissipation in the strength of the friction force. To do this, we evaluate the integral in Eq. (\ref{Quasi friction force}) numerically.
Figure \ref{Figlosses} depicts the numerically calculated ground-state friction force for atoms with vertical (left panel) and circular (right panel) polarization and different values of the metal damping rate $\Gamma_c$. It is seen that an increase in damping factor increases the friction force in the low velocity regime, whereas for large velocities the opposite trend is observed. Note that in the limit $v \to 0$, where the Born-Markov approximation breaks down, the force remains exponentially weak, albeit with a different scaling law compared to the dissipationless case \cite{Intravaia2016b}. The behavior is qualitatively similar for both polarizations. The transition between the two regimes occurs for a velocity on the order of $v_{\rm{opt}}/2$, with $v_{\rm{opt}}$  the velocity that maximizes the ground-state friction force.

It is interesting to estimate the strength of the quantum friction force. Considering the setup of Fig. \ref{Figlosses}b and a qubit with $
\gamma  = 5D = 1.6 \times 10^{ - 29} C \cdot m
$  standing at a distance of $
d = 0.1c/\omega _{{\rm{sp}}} \sim 7.3nm$  above a silver slab, the frictional force is on the order of $
0.3\,{\rm{fN}}$ for $v=0.06 c$.

\begin{figure}[h!]
\includegraphics[scale=0.58]{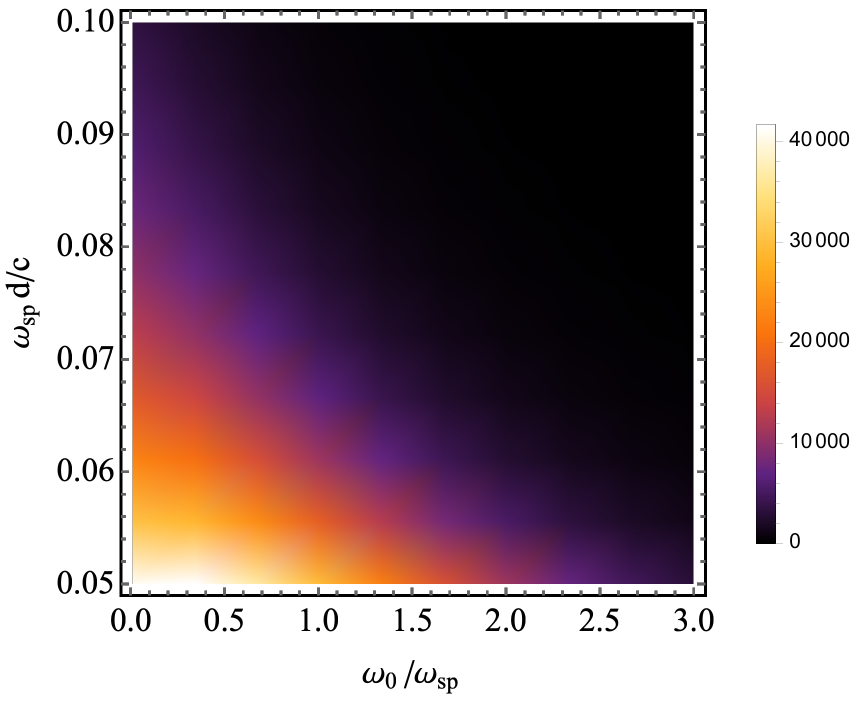}
\caption{Density plot of the normalized ground-state friction force as a function of dimensionless parameters $d \, \omega_{\rm{sp}}/c$ and $\omega_{0}/ \omega_{\rm{sp}}$ for an atom with circular polarization such that $\boldsymbol{\gamma }\sim \frac{1}{{\sqrt 2 }}\left( {{\bf{\hat x}} - i{\bf{\hat z}}} \right)$. The damping rate is $\Gamma_c = 0.2 \omega_{\rm{sp}}$ and the velocity is $v=0.05 c$. 
}
\label{Fig10Densitycirwl}
\end{figure}

Figure \ref{Fig10Densitycirwl} presents a parametric study of the intensity of the ground-state friction force as a function of $\omega_0$ and $d$ for a fixed $\omega_{\rm{sp}}$, for a level of loss $\Gamma_c = 0.2 \omega_{\rm{sp}}$, and velocity $v=0.05 c$. The transition dipole moment is assumed to be circularly-polarized. As expected, the friction force is stronger for shorter distances. 

The density plot in Fig. \ref{Fig10Densitycirwl} also reveals that the ground-state friction force is peaked for an atomic transition frequency $\omega_0 / \omega_{\rm{sp}} \approx 0$. This behavior can be understood noting that the ground-state force is mediated by plasmons with wave number ${k_x} = {k_{{\rm{P}} - }} =  - \left( {{\omega _0} + {\omega _{{\rm{sp}}}}} \right)/v$, and so the confinement of the plasmons is minimized for $\omega_0 \to 0 $. Thus, for small $v$, the interaction between the plasmons and the atom is enhanced when $\omega_0 \ll \omega_{\rm{sp}}$.

More generally, using the asymptotic expansion of the modified Bessel functions for large arguments, it is simple to check that the ground-state force (Eq. \ref{friction_lossless} with $\mathcal{P}_e=0$) depends on the transition frequency $\omega_0$ as 
${{\mathcal{F}}_x^{QS} } \sim {u^{5/2}}{e^{ - 2u}}$ with $u=\left| {{k_{{\rm{P}} - }}} \right|d$. This function reaches its maximum at $u=5/4$. Using $ {k_{{\rm{P}} - }} =  - \left( {{\omega _0} + {\omega _{{\rm{sp}}}}} \right)/v$ one finds that the $\omega_0$ that maximizes the ground-state force in the weak dissipation limit is ${\omega _{0,{\rm{opt}}}} \approx \max \left\{ {0,\frac{5}{4}\frac{v}{d} - {\omega _{{\rm{sp}}}}} \right\}$. Thus, while for small $v/d$ the maximum is reached for $\omega_0 \approx 0$ as in Fig.  \ref{Fig10Densitycirwl}, for large $v/d$ the maximum is reached for a finite frequency (not shown).  

\section{Conclusion}

In summary, using a non-relativistic formalism based on a modal expansion and the Born-Markov approximation, we derived explicit formulas for the friction force acting on a two-level system that moves with some velocity $v$ with respect to a generic material substrate. The friction force is expressed in terms of the system's Green's function and is consistent with previously reported theories based on the fluctuation-dissipation theorem. Furthermore, for the particular case of a metal described by a Drude dispersion model we employed a quasi-static approximation and derived analytical formulas for the quantum friction force in the weak dissipation limit. In particular, we have shown that the ground-state friction force is maximized for a velocity on the order of $v_{\rm{opt}} = \frac{4}{7}\left( {{\omega _0} + {\omega _{{\rm{sp}}}}} \right)d$ (in agreement with Ref. \cite{Intravaia2016b}) and for an atomic transition frequency such that ${\omega _{0,{\rm{opt}}}} \approx \max \left\{ {0,\frac{5}{4}\frac{v}{d} - {\omega _{{\rm{sp}}}}} \right\}$.

Our theory highlights the role of the transition dipole moment polarization in the frictional force, showing that atomic transitions associated with chiral-type polarizations may give the dominant contribution to the friction force in a realistic (multi-level) atom. 
Furthermore, we have shown that counterintuitively the atomic transitions that boost the friction force are associated with dipoles with an handedness which is opposite to that expected from naive classical considerations based on the spin-momentum locking. We have demonstrated that this property is a consequence of the fact that the friction force arises from interactions between positive and negative frequency oscillators due to the Doppler effect.
In typical atomic systems constrained by time-reversal symmetry, chiral transitions occur in pairs with opposite handedness. In such cases, the energy structure of the atomic system resembles a V shape, and the chiral transition that maximizes the frictional force becomes the relevant contribution. To conclude, we point out that our approach can be easily generalized to different plasmonic platforms, e.g., Weyl semimetals \cite{Weyl1,Weyl2} or materials with drift currents \cite{Shapiro, VolokotinPRL}, even when the nonlocal effects are considered.

\begin{acknowledgments} 
This work was supported in part by the Institution of Engineering and Technology (IET), by
the Simons Foundation, and by FCT/MECI through national funds and when applicable co-funded EU funds under UID/50008: Instituto de Telecomunicações. M. I. S. acknowledges the support
from  Welch Foundation (A-1261), the DARPA PhENOM program, the Air Force Office of Scientific Research (Award No. FA9550-20-1-0366), and the National Science Foundation (Grant No. PHY-2013771),  U.S Department of Energy under Award Number DE-SC-0023103, DE-SC0024882, and FWP-ERW7011.

\end{acknowledgments}

\appendix
\section{Positive and negative frequency components of the Green's function}
\label{Positive and negative frequency componenets of the Green's function}
Let $\overline {\bf{G}} \left( {{\bf{r}},{{\bf{r}}_0};\omega } \right)$ be the system's Green's function in the laboratory frame. The Green's function is defined as in Ref. \cite{Mario2018}. In conservative systems the Green's function can be expanded into electromagnetic modes. Specifically, $ -i \omega \overline {\bf{G}}$ can be written as $ - i\omega \overline {\bf{G}}  = {{\mathcal G}^ + } + {{\mathcal G}^ - } - {\bf{M}}_\infty ^{ - 1}\delta \left( {{\bf{r}} - {{\bf{r}}_0}} \right)$ \cite{Mario2018, ModalExpansions}. Here, $\bf{M}_\infty$ is the material matrix corresponding to the response of the electromagnetic vacuum (see Refs. \cite{Mario2018, ModalExpansions} for the details), and ${\mathcal G}^ \pm $ are the positive and negative frequency components of the Green's function: 
\begin{eqnarray}
\mathcal{G^+}&=&\sum_{\omega_{n\bold{k}}>0} \frac{\omega_{n\bold{k}}}{2}\frac{1}{\omega_{n\bold{k}}-\omega}\bold{F}_{n\bold{k}} (\bold{r}) \otimes \bold{F}^*_{n\bold{k}} (\bold{r}_0), \nonumber \\
\mathcal{G^-}&=&\sum_{\omega_{n\bold{k}}>0} \frac{\omega_{n\bold{k}}}{2}\frac{1}{\omega_{n\bold{k}}+\omega}\bold{F}^*_{n\bold{k}} (\bold{r}) \otimes \bold{F}_{n\bold{k}} (\bold{r}_0),  \label{greena1}
\end{eqnarray}
with poles in the ${\rm Re}\{ \omega\}>0$ and  ${\rm Re}\{ \omega\}<0$ semi-spaces, respectively.

It is useful to write ${\mathcal G}^- $ in terms of the full Green's function ${\mathcal G}= {\mathcal G}^+ +{\mathcal G}^- $. To this end, we use the fact that $\mathcal{G^{-}}$ is analytic for ${\rm Re }\{\omega\}>0$. Integrating $\frac{1}{{{\omega _0} - \omega }} \mathcal{G^-} \left( \omega  \right)$ over the complex imaginary axis and using Cauchy's (residue) theorem it follows that: 
\begin{eqnarray}
\mathcal{G^-}(\bold{r},\bold{r}_0,\omega_0)\vert_{{\rm Re}\{\omega_0\}>0}=\frac{1}{2\pi}\int_{-\infty}^{\infty} d\xi \frac{ \mathcal{G^-}(\bold{r},\bold{r}_0,i\xi)}{\omega_{0}-i\xi}. \label{greena2}
\end{eqnarray}
On the other hand, as the function $\frac{1}{{{\omega^* _0} + \omega }} \mathcal{G^-}$ is analytic for ${\rm Re }\{\omega\}>0$, the Cauchy's theorem implies that $0 = \frac{1}{{2\pi }}\int\limits_{ - \infty }^\infty  {d\xi \frac{1}{{\omega _0^* + i\xi }}} {{\cal G}^ - }\left( {i\xi } \right)$. 
It is evident from Eq. (\ref{greena1}) that $\mathcal{G^-}(\bold{r},\bold{r}_0,\omega)=[\mathcal{G^+}(\bold{r},\bold{r}_0,-\omega^*)]^*$, so that $\mathcal{G^-}(\bold{r},\bold{r}_0,i\xi)=[\mathcal{G^+}(\bold{r},\bold{r}_0,i\xi)]^*$. Therefore,
\begin{eqnarray}
0=\frac{1}{2\pi}\int_{-\infty}^{\infty} d\xi \frac{1}{\omega_{0}^*+i\xi} [\mathcal{G^+}(\bold{r},\bold{r}_0,i\xi)]^*. \label{greena3}
\end{eqnarray}
Taking the complex conjugate of Eq. (\ref{greena3}) and combining with Eq. (\ref{greena2}), we obtain
\begin{eqnarray}
\mathcal{G^-}(\bold{r},\bold{r}_0,\omega_0)\vert_{{\rm Re}\{\omega_0\}>0}=\frac{1}{2\pi}\int_{-\infty}^{\infty} d\xi \frac{ \mathcal{G}(\bold{r},\bold{r}_0,i\xi)}{\omega_{0}-i\xi}, \label{greena4}
\end{eqnarray}
where $\mathcal{G}=\mathcal{G^-}+\mathcal{G^+}$.

Using again $\mathcal{G^+}(\bold{r},\bold{r}_0,\omega)=[\mathcal{G^-}(\bold{r},\bold{r}_0,-\omega^*)]^*$ and noting that $\mathcal{G}(\bold{r},\bold{r}_0,i\xi)$ is real-valued, one sees that Eq. (\ref{greena4}) implies that:
\begin{eqnarray}
\mathcal{G^+}(\bold{r},\bold{r}_0,\omega_0)\vert_{{\rm Re}\{\omega_0\}<0}=\frac{-1}{2\pi}\int_{-\infty}^{\infty} d\xi \frac{ \mathcal{G}(\bold{r},\bold{r}_0,i\xi)}{\omega_{0}-i\xi}. \label{greena5}
\end{eqnarray}
Hence, since $\mathcal{G^-}(\bold{r},\bold{r}_0,\omega_0)\vert_{{\rm Re}\{\omega_0\}<0}=\mathcal{G}(\bold{r},\bold{r}_0,\omega_0)-\mathcal{G^+}(\bold{r},\bold{r}_0,\omega_0)\vert_{{\rm Re}\{\omega_0\}<0}$, it follows that for an arbitrary $\omega$ (not lying on the imaginary axis)
\begin{eqnarray}
\mathcal{G^-}(\bold{r},\bold{r}_0,\omega)&=&u(-\omega')\mathcal{G}(\bold{r},\bold{r}_0,\omega) \nonumber \\
&+&\frac{1}{2\pi}\int_{-\infty}^{\infty} d\xi \frac{1}{\omega-i\xi}\mathcal{G}(\bold{r},\bold{r}_0,i\xi), \label{greena6}
\end{eqnarray}
where $\omega'={\rm Re}\{\omega\}$ and $u$ is the unit step function. The subscript ``0'' of $\omega_0$ was dropped to simplify the notations.
Similarly, it is possible to write
\begin{eqnarray}
\mathcal{G^+}(\bold{r},\bold{r}_0,\omega)&=&u(\omega')\mathcal{G}(\bold{r},\bold{r}_0,\omega) \nonumber \\
&-&\frac{1}{2\pi}\int_{-\infty}^{\infty} d\xi \frac{1}{\omega-i\xi}\mathcal{G}(\bold{r},\bold{r}_0,i\xi). \label{greena7}
\end{eqnarray}
Equations (\ref{greena6})-(\ref{greena7}) give the positive and negative frequency components of the Green's function in terms of the full Green's function.

In order to extend the formulas to the case of  dissipative systems, we need to ensure that the integration path is fully contained in the upper-half frequency plane (UHP). This can be easily done noting that since $\mathcal{G}^T (\bold{r},\bold{r}_0,\omega)=\mathcal{G}(\bold{r}_0,\bold{r},-\omega)$, one can express $\mathcal{G^\pm}$ as:
\begin{eqnarray}
\mathcal{G}^{\pm}(\bold{r},\bold{r}_0,\omega)&=&u(\pm \omega')\mathcal{G}(\bold{r},\bold{r}_0,\omega) \mp  \frac{1}{2\pi}\int_{0^+}^{\infty} d\xi \nonumber \\ &&
\left[ \frac{ \mathcal{G}(\bold{r},\bold{r}_0,i\xi)}{\omega-i\xi}+ \frac{ \mathcal{G}^T(\bold{r}_0,\bold{r},i\xi)}{\omega+i\xi}\right]. \label{greena19}
\end{eqnarray}

\section{Friction force as a function of the system Green's function}
\label{AppendixGreenFriction}

In this Appendix, we write the friction force as a function of the system Green's function.

To begin with, we consider a Fourier decomposition of the functions $\mathcal{G}^{\pm}$ introduced in Appendix \ref{Positive and negative frequency componenets of the Green's function}, such that  
\begin{eqnarray}
{{\cal G}^ \pm } = \frac{1}{{2\pi }}\int {{\cal G}_{{k_x}}^ \pm }  d{k_x}.
\end{eqnarray}
Here, it is implicit that the system is invariant to translations in the $xoy$ plane, so that $\mathcal{G}^{\pm}_{k_x}$ depends on $x$ and $x_0$ as $e^{ik_x (x-x_0)}$. 
Thus, $\mathcal{G}^{\pm}_{k_x} e^{-ik_x (x-x_0)}$ is the Fourier transform of $\mathcal{G}^{\pm}$ in the $x-x_0$ variable.
Evidently,  Eq. (\ref{greena19}) implies that
\begin{eqnarray}
\mathcal{G}^{\pm}_{k_x}(\bold{r},\bold{r}_0,\omega)&=&u(\pm \omega')\mathcal{G}_{k_x}(\bold{r},\bold{r}_0,\omega) \mp  \frac{1}{2\pi}\int_{0^+}^{\infty} d\xi \nonumber \\ &&
\left[ \frac{ \mathcal{G}_{k_x}(\bold{r},\bold{r}_0,i\xi)}{\omega-i\xi}+ \frac{ \mathcal{G}_{-k_x}^T(\bold{r}_0,\bold{r},i\xi)}{\omega+i\xi}\right], \nonumber \\ \label{greena20}
\end{eqnarray}
where $ \mathcal{G}_{k_x}= \mathcal{G}_{k_x}^+ +  \mathcal{G}_{k_x}^-$. We took into account that the Green's function only depends on the relative difference ($x-x_0$) between $x$ and $x_0$.

The friction force is written in terms of the functions ${\mathcal G}_v^ \pm$ defined by Eq. (\ref{greena}).
By comparing Eq. (\ref{greena}) with Eq. (\ref{greena1}) one sees that:
\begin{eqnarray}
{\mathcal G}_v^ \pm \left( {{\bf{r}},{{\bf{r}}_0},\omega } \right) = \frac{1}{{2\pi }}\int {{\mathcal G}_{{k_x}}^ \pm \left( {{\bf{r}},{{\bf{r}}_0},\omega  + {k_x}v} \right)} d{k_x}.
\end{eqnarray}
Then, with the help of Eq. (\ref{greena20}) one finds that for positive $v$:
\begin{eqnarray}
{\cal G}_v^ + \left( {\omega _0} \right) &=&  - {I_{{\rm{nr}}}} + \frac{1}{{2\pi }}\int\limits_{ - {{\omega}_0}/v }^{ +\infty } {d{k_x}} {\cal G}_{{k_x}}^{}\left( {{\bf{r}},{{\bf{r}}_0},{\omega _0} + {k_x}v} \right), \nonumber \\
{\cal G}_v^ - \left( {\omega _0} \right) &=& {I_{{\rm{nr}}}} + \frac{1}{{2\pi }}\int\limits_{-\infty}^{ -{{\omega}_0}/v } {d{k_x}} {\cal G}_{{k_x}}^{}\left( {{\bf{r}},{{\bf{r}}_0},{\omega _0} + {k_x}v} \right), \nonumber \\
\label{Gv_final}
\end{eqnarray}
where
\begin{eqnarray}
{I_{{\rm{nr}}}} &=& \frac{1}{{{{\left( {2\pi } \right)}^2}}}\int {d{k_x}} \int\limits_{{0^ + }}^\infty d\xi \nonumber \\
&&{ \left( {\frac{{\cal G}_{{k_x}}^{}\left( {{\bf{r}},{{\bf{r}}_0},i\xi } \right)}{{{\omega _0} + {k_x}v - i\xi }} + \frac{{\cal G}_{{-k_x}}^{T}\left( {{\bf{r}}_0,{{\bf{r}}},i\xi } \right)}{{{\omega _0} + {k_x}v + i\xi }}} \right)} . \nonumber \\
\end{eqnarray}
Substituting the above formulas into Eq. (\ref{friction_modal}), one finds that the optical force satisfies ($m=x,y,z$):
\begin{eqnarray}
 \mathcal{F}_m   &=& \mathcal{P}_e (t)  \mathcal{F}_e 
+\left(1- \mathcal{P}_e (t) \right)  \mathcal{F}_g
\label{frictionA1}
\end{eqnarray}
where $\mathcal{F}_i =2 \left\{ \mathcal{J}^r_i \mp \mathcal{J}^{nr}\right\}$ ($i=e,g$), is written in terms of resonant ($ \mathcal{J}^r$) and non-resonant ($ \mathcal{J}^{nr}$) terms defined by:
\begin{eqnarray}
\mathcal{J}_{e}^r &=& \frac{\rm Re}{2\pi}   \int_{-\omega_0/v}^{+\infty}  dk_x\left\{ \boldsymbol{\tilde{\gamma}}^*\cdot \partial_m \mathcal{G}_{k_x}  (\bold{r}_0, \bold{r}_0, \omega_0+k_x v)\cdot\boldsymbol{\tilde{\gamma}}\right\}, \nonumber \\
\mathcal{J}_{g}^r &=& \frac{\rm Re}{2\pi}   \int_{-\infty}^{-\omega_0/v} dk_x\left\{ \boldsymbol{\tilde{\gamma}}^* \cdot \partial_m\mathcal{G}_{k_x} (\bold{r}_0, \bold{r}_0, \omega_0+k_x v) \cdot\boldsymbol{\tilde{\gamma}}\right\}, \nonumber \\
\label{frictionA2}
\end{eqnarray}
\begin{eqnarray}
{\cal J}_{}^{nr} = \frac{{{\mathop{\rm Re}\nolimits} }}{{{{\left( {2\pi } \right)}^2}}}\int {d{k_x}} \int\limits_{{0^ + }}^\infty  {d\xi \left[ {\frac{{{{{\bf{\tilde \gamma }}}^*} \cdot {\partial _i}{\cal G}_{{k_x}}^{}\left( {{\bf{r}},{{\bf{r}}_0},i\xi } \right) \cdot {\bf{\tilde \gamma }}}}{{{\omega _0} + {k_x}v - i\xi }}} \right.} \nonumber \\
{\rm{          }} + {\left. {\frac{{{{{\bf{\tilde \gamma }}}^*} \cdot {\partial _i}{\cal G}_{ - {k_x}}^T\left( {{{\bf{r}}_0},{\bf{r}},i\xi } \right) \cdot {\bf{\tilde \gamma }}}}{{{\omega _0} + {k_x}v + i\xi }}} \right]_{{\bf{r}} = {{\bf{r}}_0}}}
\label{frictionA3}
\end{eqnarray}

In this study, we focus on the lateral component of the force collinear with the direction of the relative motion ($x$-direction). For $m=x$, we can use ${\partial _m} = i k_x $ in the previous formulas. Furthermore, as ${\cal G}\left( {{\bf{r}},{{\bf{r}}_0},i\xi } \right)$ is real-valued, it is simple to check that ${\cal G}_{{k_x}}^{}\left( {{{\bf{r}}_0},{{\bf{r}}_0},i\xi } \right) = {\cal G}_{ - {k_x}}^*\left( {{{\bf{r}}_0},{{\bf{r}}_0},i\xi } \right)$. This property implies that the matrix 
\begin{eqnarray}
\frac{{{\cal G}_{{k_x}}^{}\left( {{{\bf{r}}_0},{{\bf{r}}_0},i\xi } \right)}}{{{\omega _0} + {k_x}v - i\xi }} + \frac{{{\cal G}_{ - {k_x}}^T\left( {{{\bf{r}}_0},{{\bf{r}}_0},i\xi } \right)}}{{{\omega _0} + {k_x}v + i\xi }} \nonumber
\end{eqnarray}
is Hermitian-symmetric. Consequently, the integrand of Eq. (\ref{frictionA3}) is purely imaginary, and thereby the non-resonant term ${\cal J}_{}^{nr}$ vanishes. Thus, the lateral force is determined exclusively by the resonant terms. This observation yields Eq. (\ref{friction force}) of the main text. Note that the non-resonant term ${\cal J}_{}^{nr}$ typically contributes to the vertical (Casimir-type) force.

\section{Atomic transition rates}
\label{Derivation of Decay Rates}
The decay rates derived in \cite{Lannebere2017} can be written in terms of the Green's functions $ \mathcal{G^{\pm}_\nu}$ defined in the main text [Eq. (\ref{greena})] as follows:
\begin{eqnarray}
\Gamma^{\pm}= \pm \frac{2}{\hbar}{\rm Im}\left\{\tilde{\gamma}^* \cdot \mathcal{G^{\pm}_\nu} {\left( {{{\bf{r}}_0},{{\bf{r}}_0},{\omega _0} + {0^ + }i} \right)} \cdot  \tilde{\gamma}\right\}.
\label{green spont.}
\end{eqnarray}
Using now Eq. (\ref{Gv_final}) one can express the decay rates in terms of the system Green's function as:
\begin{eqnarray}
\Gamma^{+}&=& \frac{1}{\pi\hbar}   \int_{-\omega_0/v}^{+\infty}  dk_x  {\rm Im}\left\{ \boldsymbol{\tilde{\gamma}}^*\cdot \mathcal{G}_{k_x}  (\bold{r}_0, \bold{r}_0, \omega_0+k_x v)\cdot\boldsymbol{\tilde{\gamma}}\right\}, \nonumber \\
\Gamma^{-}&=& \frac{-1}{\pi\hbar}   \int_{-\infty}^{-\omega_0/v}  dk_x  {\rm Im}\left\{ \boldsymbol{\tilde{\gamma}}^*\cdot \mathcal{G}_{k_x}  (\bold{r}_0, \bold{r}_0, \omega_0+k_x v)\cdot\boldsymbol{\tilde{\gamma}}\right\}. \nonumber \\
\label{decay rates11}
\end{eqnarray}
We took into account that $I_{\rm{nr}}$ evaluated for ${\bf{r}}={\bf{r}}_0$ is an Hermitian tensor and hence does not contribute to the decay rates. It is implicit that $\omega_0$ has a small positive imaginary component.

Under the quasi-static approximation (see Eq. \ref{Fourier Quasi Static}) Eq. (\ref{decay rates11}) simplifies to
\begin{eqnarray}
\Gamma^{+}&=& \frac{1}{4\pi^2\epsilon_0\hbar} \int_{-\infty}^{+\infty} dk_y  \int_{-\omega_0/v}^{+\infty}  dk_x e^{-2k_{\vert \vert}d}A(k_x,k_y) \nonumber \\
&&  {\rm Im}\{-R\left(\omega_0 + k_x v\right)\}, \nonumber \\
\Gamma^{-}&=& \frac{-1}{4\pi^2\epsilon_0\hbar} \int_{-\infty}^{+\infty} dk_y  \int_{-\infty}^{-\omega_0/v}  dk_x e^{-2k_{\vert \vert}d}A(k_x,k_y) \nonumber \\
&&  {\rm Im}\{-R\left(\omega_0 + k_x v\right)\},
\label{decay rates2}
\end{eqnarray}
with $A(k_x,k_y)$ defined as in the main text [Eq. \ref{Apolfactor}].
%

\end{document}